\documentclass[11pt]{article}

\usepackage{graphics, graphicx,epsfig, subfigure, rotating}
\usepackage{amsmath,amsfonts,latexsym,amssymb,amsthm,graphicx} 
\usepackage{palatino}
\usepackage{array, enumerate}
\usepackage{setspace}
\usepackage[usenames,dvipsnames,svgnames,table]{xcolor}
\usepackage{float}
\usepackage{titlesec}

\usepackage{natbib}
\bibpunct{(}{)}{;}{a}{,}{,}
\renewcommand{\caption}[1]{\flushleft \singlespace\hangcaption{#1}}

\title {The puzzle of partial resource use by a parasitoid wasp}
\author {Kathryn Montovan, Christelle Couchoux, Laura E. Jones,  \\H. Kern Reeve, Saskya van Nouhuys}

\begin{document}

\begin{titlepage}
{\parindent0pt 

{\linespread{1}\selectfont

\textbf{The puzzle of partial resource use by a parasitoid wasp}\vspace{.7em}

Kathryn J. Montovan$^{1,2}$, Christelle Couchoux$^3$, Laura E. Jones$^{1,4}$, H. Kern Reeve$^5$, Saskya van Nouhuys$^{3,4}$\vspace{.7em}

$^1$Center for Applied Math, Cornell University, Ithaca, New York 14853, USA\\
$^2$ Current address: Bennington College, One College Drive, Bennington, VT 05201, USA\\
$^3$ Department of Biosciences, PO Box 65 (Viikinkaari 1), University of Helsinki, 00014, Helsinki, Finland\\
$^4$ Department of Ecology and Evolutionary Biology, Corson Hall, Cornell University, Ithaca New York, 14853, USA\\
$^5$Department of Neurobiology and Behavior, Mudd Hall, Cornell University, Ithaca, New York 14853, USA\\

\textbf{Emails:} Kathryn J. Montovan: kmontovan@bennington.edu, Christelle Couchoux: christelle.couchoux@gmail.com, Laura E. Jones: lej4@cornell.edu, H. Kern Reeve: hkr1@cornell.edu, Saskya van Nouhuys: saskya@cornell.edu
}

{\linespread{2}\selectfont

\textbf{Manuscript type:} Article\\

\textbf{Elements in expanded online edition:} Appendix A - General Experimental Procedures, Appendix B -  Measuring the fitness cost of living in a highly parasitized host nest,  Appendix C - Modeling how reliably wasps avoid superparasitism, Appendix D - Probing efficiency
}}
\end{titlepage}
\pagebreak
\section*{Abstract}
% should be less than 200 words
When there is conspicuous under-exploitation of a limited resource it is worth asking, what mechanisms allow presumably valuable resources to be left unused? Evolutionary biologists have generated a wide range of hypotheses to explain this, ranging from interdemic group selection to selfishly prudent individual restraint. We consider a situation in which, in spite of high intraspecific competition, individuals leave most of a key resource unexploited. The parasitic wasp that does this finds virtually all host egg clusters in a landscape, but parasitizes only about a third of the eggs in each, and then leaves a deterrent mark around the cluster.  We first test, and reject, a series of system-specific simple constraints that might limit full host exploitation, such as asynchronous maturation of host eggs.  We then consider classical hypotheses for the evolution of restraint. Prudent predation and bet-hedging fail as explanations because the wasp lives as a large well-mixed population. Also, we find no individual benefits to the parasitoid of developing in a sparsely parasitized host nest. However an optimal foraging model, including empirically measured costs of superparasitism and hyperparasitism, can explain through individual selection, both the consistently low rate of parasitism and marking.
%\pagebreak
	
\section*{Introduction} 

Under strong resource competition, a limiting resource is predicted to  become entirely depleted. However, because of population level responses to resource availability, this does not generally occur, especially in persistent predator-prey or host-parasite interactions \citep{Abrams_2000, hassell2000host}. Here we consider an animal that, at an individual rather than population level, consistently does not deplete an apparently available resource. We examine the consistently low resource use by a parasitoid, \textit{Hyposoter horticola} (Hymenoptera: Ichneumonidae). This wasp parasitizes the butterfly \textit{Melitaea cinxia} (Lepidoptera: Nymphalidae) in \AA land, Finland. It locates host egg clusters in the landscape during the weeks before they are ready to be parasitized \citep{vanNouhuysEhrnsten04}, and monitors the egg clusters, using memorized visual landmarks \citep{vanNouhuysKaartinen08}.  The wasp parasitizes a portion of nearly every host egg cluster in the landscape, with the great majority of the parasitism in each cluster due to one female (Couchoux et al., unpublished manuscript  \textit{a}). This behavior leads to a uniform rate of parasitism, largely independent of scale of observation and host density \citep{vanNouhuysHanski02}. Here we address why individual \textit{H. horticola}, which are clearly resource limited, parasitize just a fraction of the hosts available to them, mark the clusters they parasitize, and are deterred by the markings left by others. 

We use a combination of empirical and theoretical methods to assess nine mechanisms that could potentially lead to a rate of parasitism that is systematically low. Aspects of this topic have been addressed for parasitoid wasps empirically \citep{CroninStrong93b, CroninStrong93a,  Bouskila_etal95} and theoretically \citep{AyalGreen1993, RosenheimMangel94, DriessenBernstein1999}. Here we present a broad integrated analysis in one empirical research system. The interaction between \textit{M. cinxia} and \textit{H. horticola} is especially suitable for this study because it is simple, with a parasitoid supported by a single host species. Additionally, the population and behavioral ecology for both the host \citep{Hanski2011, Ojanen_etal_2013} and parasitoid \citep{vanNouhuysHanski02, vanNouhuysKaartinen08} have been well studied on a large spatial scale. 

\subsection*{Research System} The host butterfly \textit{M. cinxia} has a Eurasian distribution. In the \AA land islands of Finland it lives as a metapopulation in a network of 4000 small meadows over an area of 3500 km$^2$. The meadows are surveyed annually, with three to five hundred of them occupied by the butterfly in any given year \citep{Ojanen_etal_2013}. Individual butterflies lay clusters of eggs on the host plants (\textit{Plantago lanceolata} and \textit{Veronica spicata}: Plantaginaceae) in June \citep{Kuussaari_etal2004}. The eggs take two to three weeks to develop, then, shortly before hatching, essentially all of the egg clusters are parasitized by \textit{H. horticola} \citep{vanNouhuysHanski02, vanNouhuysEhrnsten04}. The wasp is solitary and mobile, foraging on a larger scale than does the host \citep{Kankare_etal05}. It has no hosts other than \textit{M. cinxia} \citep{Shaw_etal2009}. Typically females spend 20 to 60 minutes parasitizing a host egg cluster \citep{Couchoux_vanNouhuys_development}, ovipositing in about a third of the eggs (field conditions: N = $43$,  $\bar{x}=0.31$  $(\pm 0.12$ SD); laboratory: N = $10$,  $\bar{x}=0.32$ $(\pm 0.07$ SD); comparison of field to laboratory conditions using Welch's t-test: $t_{\rm 22.7 }= 0.279,$  $p = 0.393$) \citep{vanNouhuysEhrnsten04}. Afterwards the wasp marks the leaves around the egg cluster, which deters conspecifics, and perhaps itself, from parasitizing the remaining hosts (Couchoux et al., unpublished manuscript \textit{a}).

\subsection*{Plausible explanations for partial resource use}
\subsubsection*{Physical limitations to parasitism} Multiple physical and physiological limitations might restrict the wasp's ability to parasitize an entire host egg cluster.  These are: {\bf Wasp egg limitation --} an individual may have few eggs available at a given time, or it may only have enough eggs to parasitize a small fraction of hosts encountered over a lifetime \citep{Bouskila_etal95, Mangel2006, Rosenheim2011}; {\bf Host egg cluster architecture --} not all of the host eggs in a cluster may be accessible to the parasitoid ovipositor \citep{Weseloh72, Hondo_etal95}; {\bf Host immune defense --} a fraction of hosts may kill the wasp eggs through immune defense \citep{LavineStrand2002};  and {\bf Ephemeral resource availability --} if host eggs develop asynchronously within a cluster, only a fraction may be susceptible while the wasp is present \citep{BriggsLatto_96}. Alternatively, if the eggs mature synchronously, while they are susceptible the wasp may only have enough time to parasitize some of them \citep{Nakamichi_et_al_2008}. Although each physical/physiological constraint could explain fractional parasitism, none would explain why a wasp applies or respects deterrent markings of the host egg clusters. 

\subsubsection*{Behavioral limitations to parasitism}  We next consider classical ecological and evolutionary scenarios that have been used to explain behavioral restraint in other resource-exploiter systems.\\

{\bf Prudent predation (parasitism).} Restrained harvesting strategies increase resource availability for future generations. This would only benefit the specific individuals practicing restraint if the species lived in small populations with limited mixing \citep{Slobodkin74, Smith64}.  Prudence has been used to explain reduced predation in some predator-prey interactions \citep{Wilson1978}. However, the \textit{M. cinxia} - \textit{H. horticola} system does not meet the requirements for this. While the host butterfly does live as networks of local populations in a fragmented landscape \citep{Hanski2011}, individual wasps are dispersive \citep{vanNouhuysHanski02}, with overlapping ranges, and only very weak geographic genetic structure \citep{Kankare_etal05}. Thus there is no opportunity for the evolution of prudence. 

{\bf  Bet-hedging.} Another possible mechanism for partial resource exploitation is distribution of reproductive effort. This can reduce variability in the expected number of surviving offspring. For instance, in temporally varying environments an organism may decrease year-to-year variation by spreading reproductive effort over multiple time periods \citep{Gillespie77,Rajonetal_2014}. While conditions do vary between years for \textit{H. horticola} \citep{HanskiMeyke05, vanNouhuys_etal2003}, an individual can only reproduce in a single season, so temporal risk spreading between years is not possible. 

In spatially structured heterogeneous environments, individuals in very small populations may increase fitness by spreading offspring over the landscape. This would decrease the probability of extinction of a particular genotype. But, in large well mixed populations there is no long term selective benefit to such reduced variance of individual success \citep{Gillespie77, Mangel2006, Hopper99}. \textit{Melitaea cinxia} larval nest mortality varies spatially \citep{HanskiMeyke05, vanNouhuys_etal2003}. However, as noted, the population of \textit{H. horticola} wasps is large and well mixed, so bet-hedging individuals would not predominate. Because we rule out both prudence and bet-hedging, neither are considered further.

{\bf  Cooperative benefits.}  
Cooperatively feeding gregarious caterpillars such as \textit{M. cinxia} rely on each other for survival \citep{Kuussaari_etal2004,Costa2006}. If parasitized caterpillars perform poorly, then the performance of highly parasitized groups would be low, decreasing individual parasitoid fitness, perhaps below the threshold necessary for the survival of the parasitoids in a host nest.  Selection due to this would favor restraint in oviposition by parasitoid females.\vspace{1.3em}

{\bf Optimal foraging, including mortality due to superparasitism and avoidance of hyperparasitism}. The final hypotheses for evolution of behavioral restraint are based on a classical optimal foraging model, wherein an individual is predicted to stop using a resource patch once the marginal benefit turns negative. For instance, the marginal value theorem predicts that individuals balance time or energy spent at a given resource patch with that spent traveling to a new resource patch \citep {Charnov1976}.  As a forager depletes a resource patch it experiences diminished returns. At some point the expected gain of leaving to find a new patch will exceed the reward of remaining, even taking into account transit time, and the forager is predicted to leave. There are many examples of consumers leaving resource patches because of diminished returns \citep{Sih_A_1980}, and this has been modeled for parasitoid wasps \citep{Wajnberg_2006, Eliassen_etal09}. We first consider a basic optimal foraging model  that assumes \textit{H. horticola} experiences diminishing returns with increased time at a host egg cluster (resource patch). The longer it stays, the more likely it is to encounter host eggs that it has already parasitized. Superparasitism is costly to solitary parasitoids \citep{RosenheimMangel94}. We measure the actual amount of superparasitism that occurs, and compare the outcome of the model when one parasitoid successfully develops within a superparasitized host, and when superparasitism causes mortality of all parasitoid eggs in a host.  We then consider the risk of  hyperparasitism (parasitism of the parasitoid), which is another potentially density dependent factor leading to diminishing returns. In order to do this realistically we use field data to measure the association of rate of hyperparasitism with rate of parasitism, and determine how it changes the outcome of the optimal foraging model.

\section*{Methods and Results}
 In the following sections we present both the experimental tests of, and results for, each potential mechanism of partial parasitism, excluding prudent parasitism and bet-hedging which were eliminated above. We start by considering the four simple biological explanations.  

\subsection*{\small Species specific biological constraint: wasp egg limitation}
Egg-limited parasitoids do not produce sufficient eggs to parasitize all of the hosts they can encounter in a patch or during a lifetime. They must thus choose which hosts to use \citep{Jervis_etal2001,Rosenheim2011}. \textit{Melitaea cinxia} egg clusters contain only about 150 eggs \citep{Saastamoinen_2007}. Couchoux and van Nouhuys (2014) found that female \textit{H. horticola} contain $\bar{x} = 550$ $(\pm 173$ SD) mature eggs in their oviducts under laboratory conditions.  Because \textit{H. horticola} is synovigenic, it is likely to mature new eggs to replace those that are used \citep{Jervis_etal2001}. A large-scale study of the genetic structure of \textit{H. horticola} in \AA land (Couchoux et al., unpublished manuscript \textit{b}) showed that, on average, a successful mother parasitizes about four egg clusters, two of which survive the winter \citep{vanNouhuys_etal2003}. So, although evolutionary pressures may have brought \textit{H. horticola} to this point, at present in \AA land the wasp is not strongly egg limited. Most individuals successfully parasitize significantly fewer hosts than they have eggs, and egg limitation cannot dictate the average foraging behavior.  Additionally, if host egg clusters differed in quality and wasps were choosey, then the rate of parasitism is predicted to vary greatly from cluster to cluster, which it does not, even with respect to egg cluster size \citep{vanNouhuysEhrnsten04,Couchoux_vanNouhuys_development}.

\subsection*{\small Species specific biological constraint: Host egg cluster architecture 
}
\textit{Melitaea cinxia} lay eggs in mounds. For some insect species the inner eggs in mounds are inaccessible to the parasitoid ovipositor \citep{Weseloh72, Hondo_etal95}, with up to half of the eggs in the protected inner layers \citep{Friedlander85}. To find out if \textit{H. horticola} is restricted to the outer eggs we compared parasitism rates of inner and outer layers of host egg clusters. Eleven egg clusters were exposed to parasitism by \textit{H. horticola} in the laboratory (see Appendix \ref{App:GeneralExpProc} for methods). Seven wasps were used, with three each parasitizing a single cluster, and the other four each parasitizing two clusters. Immediately after parasitism the outer layer of eggs was separated from the rest of the cluster. Both categories were then dissected to determine the fractions parasitized. The inner and outer eggs were parasitized equally (outer eggs $\bar{x} = 0.43$ $(\pm 0.16$ SD), inner eggs:  $\bar{x} = 0.48$ $(\pm 0.22$ SD)) (paired t-test: $t_{8}$ = 1.1929, $p = 0.2604$), indicating that mounding does not protect the inner host eggs from parasitism.

\subsection*{\small Species specific biological constraint: Host egg immunological defense }
\label{sec:HostDefense}

Insects can defend themselves against endoparasitoids by encapsulating or otherwise preventing development of parasitoid eggs or larvae \citep{LavineStrand2002}. For instance, \textit{M. cinxia} caterpillars encapsulate up to half the larvae of the parasitoid \textit{Cotesia melitaearum} (Hymenoptera: Braconidae) \citep{vanNouhuysETAL2012}. If the majority of \textit{M. cinxia} were resistant to parasitism by \textit{H. horticola}, then the low rate of successful parasitism would be explained by host immunity. However, encapsulation of \textit{H. horticola} would have to occur early in host development (before the host hatches from the egg) which is both unlikely and costly \citep{Schmid_Hempel_2005, ArdiaGantz2012}. Furthermore, there is no evidence of encapsulation.  For instance, no dead parasitoid eggs were found in \textit{M. cinxia} caterpillars dissected within hours of hatching, such as those used used in this study. These early dissected caterpillars also did not have a higher incidence of parasitism (N = $64$, $\bar{x}=0.30$  $(\pm 0.13$ SD)) than in previous studies in which the caterpillars were dissected later in development (34\%) \citep{vanNouhuysEhrnsten04}, or upon adult emergence (36\%) \citep{vanNouhuysPunju10}.

Nonetheless, we approached this idea comparatively.  Assuming resistance to parasitism is costly, hosts from places where the parasitoid occurs may have evolved resistance, whereas without the parasitoid there would be no or low resistance \citep{KRAAIJEVELD_2002}. In the laboratory we compared the rate of successful parasitism of \textit{M. cinxia} from \AA land with those from Morocco, which lacks  \textit{H. horticola}. The only known parasitoid of \textit{M. cinxia} caterpillars in Morocco is \textit{C. melitaearum} (van Nouhuys, Pers.~Obs.), which parasitizes older caterpillars \citep{vanNouhuysPunju10}. In this experiment 11 egg clusters from butterflies from \AA land and 15 from Morocco were parasitized in the laboratory, each by a different  by \textit{H. horticola} individual from \AA land. For methods see Appendix \ref{App:GeneralExpProc}. Eggs from both origins were parasitized at the same frequency ($28\%\pm 17\%$ SD, Welch's t-test $t_{19.46} = -0.0047, p = 0.9963$,  Table \ref{table:AlandEstoniaMorocco}), indicating no local resistance in \AA land.

\subsection*{\small Species specific biological constraint: Ephemeral resource availability }
Temporal asynchrony of the adult parasitoid with the susceptible stage of the host can create a short opportunity for parasitism \citep{BriggsLatto_96}. The window of time \textit{H. horticola} has to parasitize eggs within a host cluster depends on the length of time individual eggs are susceptible, and on the degree of synchrony of hatching within a cluster. 

\textit{Melitaea cinxia} eggs start out bright yellow. After $12$ to $15$ days the eggs change to a creamy color, develop dark specks, turn grey, and then just before the caterpillar hatches, the top of the egg becomes nearly black. Wasps do not probe clusters of bright yellow eggs, and once the caterpillars start to hatch, the wasps are no longer attracted to the cluster \citep{Castelo_etal2010}. In order to determine the association of developmental stage of the host eggs with the rate of parasitism, we observed which visible phases of egg development were parasitized by the wasp. 
Thirty-four host egg clusters of different stages of maturity (starting with all of the eggs creamy) were exposed to parasitism in the laboratory (Appendix \ref{App:GeneralExpProc}). Each of eleven wasps was used several times (two to seven). Immediately after parasitism, the eggs within each cluster were separated into four categories: creamy, speckled, grey topped, and black topped.  Upon hatching the caterpillars were dissected to determine which were parasitized.

We analysed the association of parasitism with egg maturity category using logistic regression in JMP \citep{JMP_2012}.  The explanatory variable was egg maturity (creamy, speckled, grey-topped, and black-topped).  Egg cluster ID and wasp ID were included in the model to account for intra-cluster or intra-individual correlation in responses. The number of replicates was too unbalanced to include observation number for each wasp in the analysis. Previous experience is known to affect wasp behavior generally \citep{Vet_et_al_1990}, however it is unlikely to influence the outcome here because, in similar experiments using \textit{H. horticola}, no change in rate of parasitism was detected after the first oviposition experience \citep{Castelo_etal2010}.  We found that parasitism differed among  wasps  ($\chi^{2}_{10} = 68.1924, p < 0.0001$) and among host egg clusters ($\chi^{2}_{8} = 55.3932, p < 0.0001$) but did not significantly differ between the four egg maturity classes ($\chi^{2}_{3} = 4.1523, p = 0.2455$). Summed over the egg clusters (each containing only some of the maturity classes), the rate of parasitism was $0.15$ creamy, $0.27$ speckled, $0.27$ grey topped, and $0.30$ black topped.

To determine the amount of time a cluster contains susceptible eggs, we took hourly photographs of ten egg clusters over the last one to five days of development, and calculated the amount of time that at least 95\% of the eggs in the cluster were in one of the last three visible stages of development (speckled, grey topped, black topped). The minimum interval of susceptibility for these ten egg clusters was approximately 28 hours, and the mean was $\bar{x} = 64$ $(\pm 38$ SD). To demonstrate the pattern of development within an egg cluster we video-taped an egg cluster over 80 hours, from when the first eggs became susceptible until hatching (Fig. \ref{fig:ClusterDevelopment}). In this case, the susceptible period lasted at least 40 hours, which is much greater than the 20 minutes to one hour spent at a cluster. Because a wasp can probe approximately one egg per minute (computed in Appendix \ref{App:CalcB}), \textit{H. horticola} is not constrained by rate or synchrony of egg development in a cluster. 

In sum, \textit{H. horticola} has enough mature eggs in its ovaries and oviducts to parasitize multiple whole clusters. All of the eggs in the cluster are physically accessible to the ovipositor, and the eggs are susceptible to parasitism for much longer than a wasp attends to the cluster. Although some other physical or physiological factor could keep a parasitoid from parasitizing more hosts, we have tested those that seem plausible. 

\subsection*{\small Cooperative benefits of unparasitized hosts 
}
\textit{Melitaea cinxia} caterpillars live in gregarious family groups until their final instar \citep{Kuussaari_etal2004}. The fitness of a cooperative group depends on the performance of each individual. Thus, if parasitized caterpillars contribute less to the group than unparasitized caterpillars, increasing the group rate of parasitism would decrease the performance of parasitoids developing within the hosts.  We determined the effect of rate of parasitism on parasitoid performance by manipulating the fraction of caterpillars parasitized  per nest in a replicated laboratory experiment, measuring the rate of development, weight at diapause, size at pupation of the host and parasitoid, and production of silk by the hosts at diapause. For methods, including the statistical models used, see Appendix \ref{App:Cooperativebenefits}.

At the prediapause stage the rate of parasitism ranged among nests from 12\% to 65\% ($\bar{x}=36$\%, $ \pm 15$\% SD) among the 30 caterpillar groups.  Parasitized caterpillars developed from second instar to diapause in $\bar{x} = 29.28 $ days $(\pm 2.82$ SD). The development time differed among replicate groups, but was unrelated to the rate of parasitism in a group ($F_{1, 24.72} = 0.0092$, $p = 0.9242$). The pre-diapause development time of unparasitized caterpillars was about the same and also did not differ with rate of parasitism of the group ($F_{1,24.72} = 0.0110$,  $p = 0.9172$).
At diapause parasitized caterpillars weighed $\bar{x}=9.47$ mg  $(\pm 5.93$ SD), which did not vary with rate of parasitism ($F _{1,36.88} = 1.9877$,  $p = 0.1670$).  Very few caterpillars died in this experiment so mortality was not analyzed. Upon molting to diapause the caterpillars produced silk to make a winter nest.   Groups with a high rate of parasitism produced the most silk ($F_{4,34} = 8.9052$, $p < 0.0001$). This effect was due to the especially high production of silk by the most parasitized groups (post-hoc test $F_{1,34} = 32.51$,  $p < 0.0001$). 

	For the post-diapause caterpillars the rate of parasitism ranged from 5\% to 61\% ($\bar{x}=32\%$ $(\pm 13\%$ SD)). There was nearly 30\% mortality due to a viral infection that came late in the experiment. The mortality of parasitized caterpillars due to the virus differed among replicate groups (maximum likelihood $\chi^{2}_{36} = 70.6167, p = 0.0005$), but was unrelated to rate of parasitism ($\chi^{2}_{1}  = 0.00013$, $p = 0.9971$).   Parasitized caterpillars developed from diapause to pupation in $\bar{x}=27.55$ ($ \pm 2.79$ SD) days, which increased marginally with rate of parasitism ($F_{1,30.5} = 3.5813$, $p = 0.0680$). There was no association of development rate of unparasitized caterpillars with parasitism rate ($F_{1,21.36} = 1.1915$, $p = 0.2872$). Parasitoid pupae weighed $\bar{x}=48.96$ mg $(\pm 12.08$ SD), and did not vary with rate of parasitism. Butterfly pupae weighed $\bar{x}=177.13$ mg $(\pm  28.18$ SD).  In contrast to the parasitoid, butterfly pupal weight decreased with increasing parasitism rate ($F_{1,26.94} = 5.5352$, $p = 0.0262$). 
 
Based on these experiments, we see no great benefit for \textit{H. horticola} of being in a nest with low parasitism. It is unlikely, but possible of course, that the one day (3\%) increase in development rate between the lowest and highest rate of parasitism could have a large negative effect over a one year lifecycle. The positive association of parasitism with silk production warrants further study, because silk is positively associated with winter nest quality which is important for overwintering success  of the host \citep{Kuussaari_etal2004}, and hence the wasp. Because of the experiment design we could measure the effects of abnormally low, but not extremely high parasitism.  Thus, our treatments safely span the normal range \citep{vanNouhuysEhrnsten04}, but do not address the possible negative effects of very high parasitism (greater than twice the normal rate).

\subsection*{\small Optimal foraging 
} 
\label{sec:OF}
Optimal foraging models are used to predict how an animal should partition limited time between procuring resources and using them \citep{Charnov_1984, Charnov_Skinner}. \textit{Hyposoter horticola} has a limited time to forage for host egg clusters distributed in a landscape and parasitize them, so an optimal foraging model seems appropriate. At a host egg cluster \textit{H. horticola} probes host eggs unsystematically, making haphazard passes across the cluster (Montovan, Pers.~Obs.). Because only one \textit{H. horticola} larva can develop within each caterpillar, foraging efficiency diminishes over time as the wasp increasingly encounters previously parasitized hosts (Fig.~\ref{fig:ParasitismEfficiency}). The wasp is predicted to ultimately leave. Other within-patch density dependent factors, such as host mortality due to superparasitism and hyperparasitism, would further decrease the marginal benefit from continuing to parasitize a host cluster.\\

{\bf\small  Observed superparasitism rates.}  For solitary parasitoid species, superparasitism results in mortality of parasitoids, and sometimes hosts. Some species are able to avoid superparasitism while others are not \citep{Godfray_etal1994}. To determine the potential cost of superparasitism, we assessed its frequency. Host egg clusters were exposed to parasitism naturally in the field (N=5) and in the laboratory (N=25), and the caterpillars were dissected upon hatching to count the number of parasitoid eggs. These dissections showed that although only one wasp reaches maturity within a given host \citep{vanNouhuysPunju10}, superparasitism occasionally occurs (Fig. \ref{fig:Superparasitism}). We used these data to estimate the probability of superparasitism (Appendix \ref{App:Superparasitism}), and found that when a wasp encounters a previously parasitized egg it successfully avoids parasitizing that egg again $77\%$ of the time (dotted line in Fig. \ref{fig:Superparasitism}). The strong avoidance of superparasitism suggests that it is costly, either due to risk associated with superparasitism or (if it had not been excluded as a possibility), egg limitation.\\

{\bf\small Optimal foraging modeling.}  
The expected number of host eggs in a cluster parasitized singly, or multiple times, is 
\begin{equation}
N_p\approx N(1-e^{{-bt}/N}),
\label{equation1}
\end{equation}
where $N$ is the total number of hosts in a cluster, $b$ is the probing rate (taken from laboratory data, Appendix \ref{App:CalcB}), and $t$ is the time spent probing the cluster. 

The parasitism frequency function (Eqn. $\ref{equation1})$ predicts the number of parasitoid offspring in a cluster. It assumes that each probe by the wasp is independent and random (Appendix \ref{App:Superparasitism}), only one wasp parasitizes each cluster (Couchoux et al., unpublished manuscript \textit{a}), and if an egg is superparasitized, exactly one wasp larva will survive (Eqn.~\ref{equation1}). We were not able to experimentally determine whether one offspring survives or all offspring die in superparasitized hosts, so we also used a model in which superparasitism kills both wasps so the number of parasitoid offspring is the expected number of host eggs parasitized exactly once:
\begin{equation}
N_1 \approx \frac{ N e^{-{bt}/N}}{z} (e^{{btz}/N}-1), 
\label{equation2}
\end{equation}
\noindent where $z$ is the probability of avoiding multiparasitism, calculated in Appendix \ref{App:Superparasitism}. The average fitness is then defined as the parasitism rate (similar to the net energy intake functions in \citet{Charnov1976}), which is the number of eggs parasitized in each cluster ($N_p$ or $N_1$) divided by the time the wasp spends 'searching for' ($t_s$) and parasitizing ($t$) a cluster. The search time $t_s$ , in its simplest form, is the time it takes a wasp to reach the next available host egg cluster. Natural selection acts upon $t$, the time spent probing each cluster. The fitness functions (generically $w(t)$) representing parasitism efficiency without mortality of multiply parasitized eggs ($w_1(t)$), and assuming complete mortality of multiply parasitized eggs ($w_2(t)$), are 
\begin{equation}
w_1(t)=\frac{N_p}{t_s+t}\approx \frac{N (1-e^{-{bt}/N})}{t_s+t}
\label{eqn:OptFor} 
\end{equation}
\begin{equation}
w_2(t)=\frac{N_1}{t_s+t}\approx \frac{ N e^{-{bt}/N} (e^{{btz}/N}-1)}{z(t_s+t)}.
\label{eqn:OptFor2} 
\end{equation}
\noindent

\noindent  To maximize the fitness ($w(t)$) with respect to time spent parasitizing, $t$, we differentiate $w(t)$ and solved for $t$ when $\frac{d w(t)}{dt}=0$, and $\frac{d^2 w(t)}{dt^2}<0$, finding the optimal value of $t$ numerically and then using this value in the expressions for $N_p$ or $N_1$. 

Figure \ref{fig:OFSolns} shows the resulting optimal fraction parasitized for both parasitism functions. Over realistic ranges of egg cluster size (N) (Fig. \ref{fig:OFSolns}A), and probing rate ($b$) (Fig. \ref{fig:OFSolns}B), the optimal fraction parasitized is fairly insensitive to changes in the number of eggs or probing rate. Because the expected time it takes a wasp to reach another host egg cluster ($t_s$) is unknown, we tested the model over a large range of searching times. For intermediate values of $t_s$ ($0.25 < t_s<1$ hr) and realistic values of $N$ and $b$, both models predict an optimal fraction parasitized close to the observed $30\%$ (Fig. \ref{fig:OFSolns}C). Including mortality due to superparasitism (bold black dashed line) lowers the optimal parasitism rates and creates a larger range of search times, $t_s$, for which we would expect to see the wasp parasitize close to $30\%$ of hosts. Thus, optimal foraging with diminishing returns due to random probing (with or without superparasitism as entirely lethal), can explain the observed parasitism frequencies if the wasp's searching time is intermediate (about a half hour).  A search time of a few minutes leads to a very low rate of parasitism of less than 30\%. Long search time ($>2$ hrs) (Fig. \ref{fig:OFSolns}D) leads to a parasitism rate above 60\%.

{{\bf\small Avoiding hyperparasitism.
} Parasitoids might also behave so as to reduce the risk of mortality of their offspring, imposed by natural enemies \citep{AyalGreen1993}. The hyperparasitoid \textit{Mesochorus stigmaticus} (Hymenoptera: Ichnuemonidae) parasitizes \textit{H. horticola} larvae within \textit{M. cinxia} caterpillars. Multiple \textit{M. stigmaticus} females visit a caterpillar nest over several weeks during the summer, spending from minutes to hours there \citep{Reichgelt:Thesis:2007}. Most host egg clusters are hyperparasitized, at a rate of up to 50\%. We empirically determined the association of rate of hyperparasitism with rate of parasitism by \textit{H. horticola}, and included this in the optimal foraging model. We also compared parasitism frequencies of  \textit{H. horticola} from populations with \textit{M. stigmaticus} (\AA land) and without (Estonia), to see if the \textit{H. horticola} from \AA land have evolved low parasitism frequency in the presence of the hyperparasitoid. 

We measured the hyperparasitism frequency over a range of parasitism frequencies using two data sets. The first was 16 field-collected naturally parasitized and hyperparasitized nests. To extend the range of parasitism rate and standardize for nest size and location, we also constructed nests of 60 \textit{M. cinxia} caterpillars, as in the experiment on cooperative benefits (Appendix \ref{App:Cooperativebenefits}). We left nests containing naturally parasitized caterpillars undiluted (N= 12), diluted 1:1 (N= 10), and diluted 2:1 (N=11), this resulted in 10\% to 60\% parasitism.  We then placed the randomized nests in ten different habitat patches to be naturally hyperparasitized by \textit{M. stigmaticus}. After three weeks in July when \textit{M. stigmaticus} was active, we retrieved the nests reared the caterpillars. The following spring we recorded the numbers that produced adult butterflies, \textit{H. horticola} or \textit{M. stigmaticus}.

We used logistic regression to estimate the relationship between the fraction of the cluster that was parasitized by \textit{H. horticola} and the probability that those parasitoid larvae were hyperparasitized. The dependent variable measured whether each parasitized host egg was also hyperparasitized by \textit{M. stigmaticus}, where $p=N_p/N$ is the fraction of parasitized host caterpillars (Fig. \ref{fig:MesData}A). The independent variable was the fraction of the cluster that was parasitized by \textit{H. horticola}. The intercept was estimated as $-1.86 $ ($SE=0.27$, $p<0.001$), and the coefficient associated with the parasitism frequency was $2.7$ ($SE=0.5$, $p<0.001$). Combining these estimates, the probability that a parasitized egg was hyperparasitized is thus fit as
\begin{align}
h(p)=1/(1+e^{1.86-2.7p}).
\label{eqn:h(p)} 
\end{align}

Under this pressure of hyperparasitism, the expected number of parasitoid offspring per cluster $N_{ps}$ is:
$$N_{ps}=N \cdot  p (1-h(p)).$$ 
$N_{ps}/N$ is shown in Fig. \ref{fig:MesData}B. This leads to a new version of the optimal foraging model with the following fitness function (using Eqn. \ref{equation1}):
\begin{align}
w_3(t)=\frac{N_{ps}}{t_s+t}&= %\frac{N \cdot  p (1-h(p))}{t_s+t}\notag \\
%&= w_1(t) \left(1-\frac{1}{1+e^{-1.86-2.7p}}\right) \notag\\
%&= 
w_1(t) \left(1-\frac{1}{1+e^{-1.86-2.7(1-e^{-{bt}/N})}}\right). 
\label{eqn:OptFor3} 
\end{align}

The same numerical methods described for the basic optimal foraging model were used to determine the optimum parasitism frequencies (dashed grey lines in Fig. \ref{fig:OFSolns}). The predicted optimal fraction parasitized is similar to that in which all parasitoid larvae in superparasitized hosts all die (black dashed lines in Fig. \ref{fig:OFSolns}).

As a second approach to the potential effects of hyperparasitism on rate of parasitism, we compared \textit{H. horticola} from \AA land with those from Estonia (250 km by sea from \AA land), which is free of  hyperparasitism (van Nouhuys, Pers.~Obs.). If \textit{H. horticola} has evolved to parasitize at a low frequency to avoid a density dependent hyperparasitism in \AA land, then we might expect individuals from Estonia not to exhibit such restraint, and to parasitize a larger fraction of the hosts in a cluster. 

In a fully crossed experiment \textit{H. horticola} from \AA land and Estonia were offered \textit{M. cinxia} eggs from \AA land and Estonia (Appendix \ref{App:GeneralExpProc}). We compared the frequency of parasitism using a generalized linear model in R \citep{R}. Parasitism rate was modeled as a function of egg cluster origin (\AA land, Estonia), wasp origin (\AA land, Estonia) and the interaction between wasp and egg origin. See table \ref{table:AlandEstoniaMorocco} for the number of replicates, and results for each treatment. On average, 36\% of the eggs in a cluster were parasitized. There was no significant difference between wasp ($t_{1,47} = 0.891$,  $p = 0.377$) or egg ($t_{1,47} = 1.763$, $p = 0.085$) origins, and no interaction between them ($t_{3,45} = 1.093$, $p = 0.280$). Thus, in spite of  evidence from the optimal foraging model that the wasp is predicted to decrease rate of parasitism to avoid hyperparasitism, this experiment does not support the hypothesis that \textit{H. horticola} from \AA land have evolved restraint because of pressure from the hyperparasitoid.

\section*{Discussion}

\textit{Hyposoter horticola} forages in a competitive environment in which virtually all host egg clusters are found, and many are monitored by multiple females until they become susceptible to parasitism. Yet only about a third of each host egg cluster is parasitized, each primarily by one female. Here we examined explanations for why the wasp does not further exploit its host, using experiments, comparative studies and mathematical modeling.

\subsection*{\small Simple biological and physiological constraints}
Host egg cluster architecture, synchrony or asynchrony in the development rate of host eggs within a cluster, or the early immune response of the host do not appear to limit the wasp. An individual female also contains more eggs than needed to parasitize a single host egg cluster, and because it does not parasitize many egg clusters in its life and can probably make new eggs as they are used, a wasp, on average, is unlikely to be egg limited over a lifetime. Because \textit{H. horticola} has specialized biology and an extremely narrow host range it is unsurprising that the wasp is not limited in these ways.  We would only expect simple biological limitations to be effective constraints if, instead, the parasitoid were poorly adapted to the host.

\subsection*{\small Behavioral restraint: Prudent parasitism, risk spreading, cooperative benefits}
When individuals are physically and physiologically able to further exploit a resource, but do not, they are exhibiting behavioral restraint.  We rejected prudence and risk-aversion as explanations because in \AA land \textit{H. horticola} has a large population that is well-mixed across the landscape (Kankare et al., 2005). We also found that wasps do not benefit from developing in host nests with low parasitism: although \textit{M. cinxia} caterpillars live gregariously and rely on cooperative behavior to survive, the fraction parasitized did not significantly affect the pre-diapause or post-diapause developmental rates, weight or survival of the wasps.  The lack of a measurable fitness-cost of parasitism is unsurprising because a parasitoid larva stays extremely small (1st instar) throughout most of the development of the host caterpillar, and then grows rapidly, consuming the entire host, just before it would have pupated \citep{vanNouhuysPunju10}. 

\subsection*{\small Behavioral restraints: Optimal foraging}
Unlike the previous scenarios, optimal foraging shows promise as an explanation for partial resource use by \textit{H. horticola}.  In the most basic model, efficiency at a host egg cluster decreases solely because the wasp probes randomly, and only one larva can develop within each host. As the wasp spends more time at the cluster it finds fewer and fewer unparasitized eggs and thus benefits from leaving the cluster to find another (Fig. \ref{fig:ParasitismEfficiency}). Such a model has been used to predict the very low rate of parasitism by \textit{Anagrus delicatus}, a tiny parasitoid of leaf hoppers \citep{RosenheimMangel94}, which is unable to distinguish between parasitized and unparasitized host eggs \citep{CroninStrong93b}. Although \textit{H. horticola} can avoid superparasitism with $77\%$ accuracy, it still experiences diminished returns as the rate of parasitism increases. If superparasitized hosts die, then there are eventually even negative returns as parasitism increases. The key general principle is that extreme under-exploitation of resources can occur when exploitation progressively reduces the value of the remaining resources in the patch \citep{Charnov1976}.  

Any other factors that add cost with increased parasitism, or time at a cluster, effect predictions of the optimal foraging model.  We hypothesized that the time at a cluster would be reduced due to density dependent hyperparasitism \citep{AyalGreen1993}. This is a compelling multitrophic behavioral explanation \citep{Volk1999} which does indeed reduce the optimal rate of parasitism. In our models,  both risk of superparasitism and hyperparasitism similarly reduced optimal rate of parasitism with respect to search time, egg cluster size (N) and probing rate (b)  (Fig. \ref{fig:OFSolns}).  Interestingly, in spite of this cost there is no evidence that wasps from \AA land have evolved a lower parasitism rate than wasps from Estonia, where there are no hyperparasitoids. 

The optimal foraging model is sensitive to the search time between clusters, $t_s$, and predicts that the rate of parasitism is about 30\% when  $t_s$ is about a half hour. On the one hand we know that \textit{H. horticola} successfully parasitizes only a few clusters in a few weeks (Couchoux et al., unpublished manuscript \textit{b}), so the $t_s = 30$ minutes is too short. On the other hand, the wasp knows the locations of the clusters ahead of time, and most travel times are only seconds to minutes \citep{vanNouhuysEhrnsten04, vanNouhuysKaartinen08}, so $t_s = 30$ minutes is too long. In order for the optimal foraging (time budget) model to be applicable to this research system we have to interpret  $t_s$ differently, taking into account activities associated with the strong intraspecific competition among foraging females  \citep{vanNouhuysEhrnsten04, Hardy_etal_2013, Couchoux_vanNouhuys_development}.  For \textit{H. horticola}, a  successful individual is one who is a strong competitor, devoting a large fraction of its time to monitoring and attending host egg clusters that are not yet ready to be parasitized.  Any time the the wasp spends parasitizing is time not spent competing, thus reducing future gain. Under this scenario $t_s$ represents time that must be invested to protect, on average, one future egg cluster.

\subsection*{\small Behavioral restraint: The role of deterrent marking}
After \textit{H. horticola} has finished parasitizing, it applies a chemical mark on the leaves around the cluster (Couchoux et al., unpublished manuscript \textit{a}). Other parasitoid species are known to mark individual hosts or clusters and modify their search behavior in response their own marking or the marks of conspecifics \citep{HollerHormann1993, BernsteinDriessen1996}. None of the simple biological explanations for consistently low parasitism could provided explanations for deterrent marking by \textit{H. horticola}. Nor could prudence, bet-hedging, or cooperative benefits. The optimal foraging model does. If a wasp leaves when additional parasitism would reduce its expected fitness (due to risk of self-superparasitism or hyperparasitism) it may benefit by leaving a mark to assist itself in avoiding further parasitizing the same cluster \citep{Mangel89, Veraldi_etal2005}. A second wasp that approaches the same cluster would also maximize its fitness by leaving to search for an unused cluster. This makes it intuitive that a wasp might both mark and respect a deterrent mark left by another wasp \citep{RoitbergMangel_1988, Hoffmeister_1997}. As an aside, just as it is adaptive for some solitary parasitoids to engage in superparasitism \citep{vanAlphen_etal_1990, Speirs_etal1991}, surely some individuals would benefit from further parasitizing a previously used cluster. We might then expect that the effectiveness of the deterrent mark decreases with increasing competition for host egg clusters.  

\subsection*{Conclusion}
Any time an individual exercises extreme restraint in the use of an apparently available yet limiting resource, we wonder why. This paper illustrates that, while there are multiple potential explanations for the evolution and maintenance of low exploitation of available resources, many turn out to be implausible. None of the simple physical or physiological mechanisms examined explain the pattern. Two well known behavioral mechanisms, prudence and bet-hedging, are also not relevant because the wasp population is large and well mixed.  We also found no indication that individuals benefit from being in a sparsely parasitized cooperatively feeding host group. The surviving candidate explanation is that \textit{H. horticola} practices partial parasitism and deterrent marking as a way to forage optimally for hosts and avoid superparasitism, with the avoidance of density dependent hyperparasitism as a further incentive for restraint.  The plausibility of the optimal foraging hypothesis depends on a $t_s$ of about a half hour, which in this system should be considered not as search time, but as the time not spent parasitizing or competing for hosts.  In this study we found, as has been found in many circumstances by others, that individual selection is a stronger force than bet-hedging or prudence through group selection, and should be carefully disentangled when thinking about the evolutionary causes of any surprisingly low resource use.  

\section*{Acknowledgements}
We thank D. Muru for help in the group fitness study, L. Salvaudon for help in the egg cluster architecture study, S. Ikonen, T. Lahtinen, M. Brunfeldt and E. Metsovouri for laboratory help, A. Ruina for help with figures and and Appendix \ref{App:Superparasitism}, and S. P. Ellner for advice on fitting multinomial data in Appendix \ref{App:Superparasitism}. T. Collet, T. Day, X. Fauvergue, I. Hanski, M. Mangel, A. Ruina, K. Woods and two anonymous reviewers made helpful comments on the manuscript. Funding came from Academy of Finland grant numbers 250444, 213547, 125553 to S. van Nouhuys, and a travel grant for K. Montovan from Cornell University Department of Ecology and Evolutionary Biology.

\section* {}

 \begin{table}[H]
\begin{center}
\footnotesize
\begin{tabular}{llcc}
\textbf{Host origin} & \textbf{Wasp origin} & $\boldsymbol{n}$ & $\boldsymbol{\mu_p\pm\sigma_p}$\\ \hline
\AA land&\AA land&11&$28\%\pm 21\%$\\ 
\AA land&Estonia&14&$35\%\pm 18\%$\\ 
Estonia&\AA land&10&$43\%\pm 17\%$\\ 
Estonia&Estonia&14&$38\%\pm18\%$\\ 
Morocco&\AA land&15&$28\%\pm18\%$\\ \hline
\end{tabular}
\normalsize
\end{center}
\caption{ \baselineskip=2\baselineskip Summary of results from the \AA land, Estonia,  and Morocco parasitism comparison studies (Appendix \ref{App:GeneralExpProc}). $n$ is the number of host clusters. $\mu_p$ is the mean fraction parasitized, and $\sigma_p$ is the standard deviation of the fractions parasitized.}
\label{table:AlandEstoniaMorocco}
\end{table}
 
\begin{figure}[H]
\centering
\includegraphics[trim=0.1cm 0.7cm 1.4cm 2.7cm, clip, width=.9\textwidth]{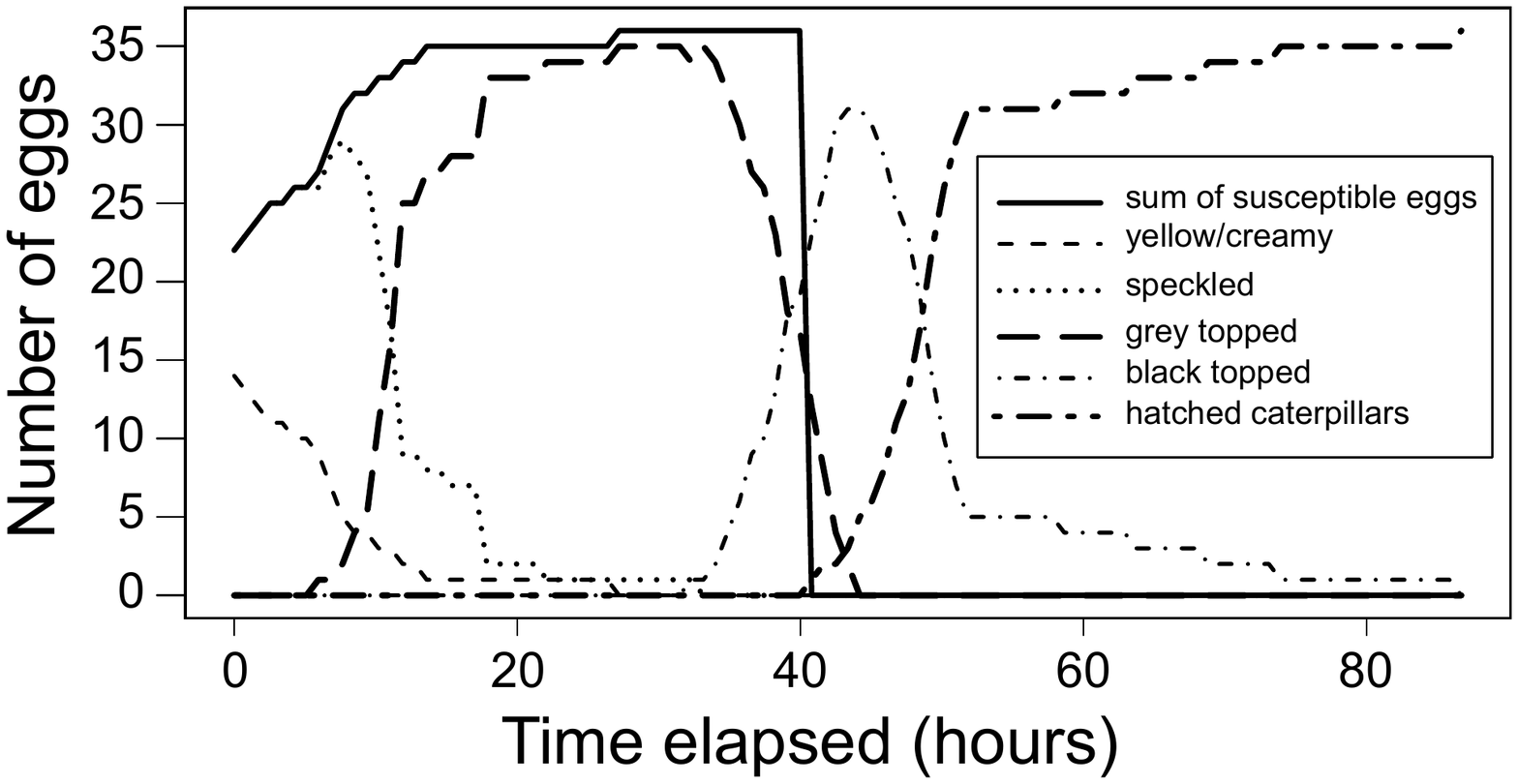} 
\caption{ \baselineskip=2\baselineskip\textbf{The temporal pattern of host egg development in one \textit{M. cinxia} egg cluster}. Lines show the number of host eggs in each developmental stage. The black solid line (sum of the susceptible egg stages) falls abruptly as the first caterpillar emerges, after which the wasp will no longer parasitize any eggs in the cluster. The susceptible time for the egg cluster is at least 40 hours.}
\label{fig:ClusterDevelopment} 
\end{figure}

\begin{figure}[H]
\centering
\includegraphics[width=.85\textwidth]{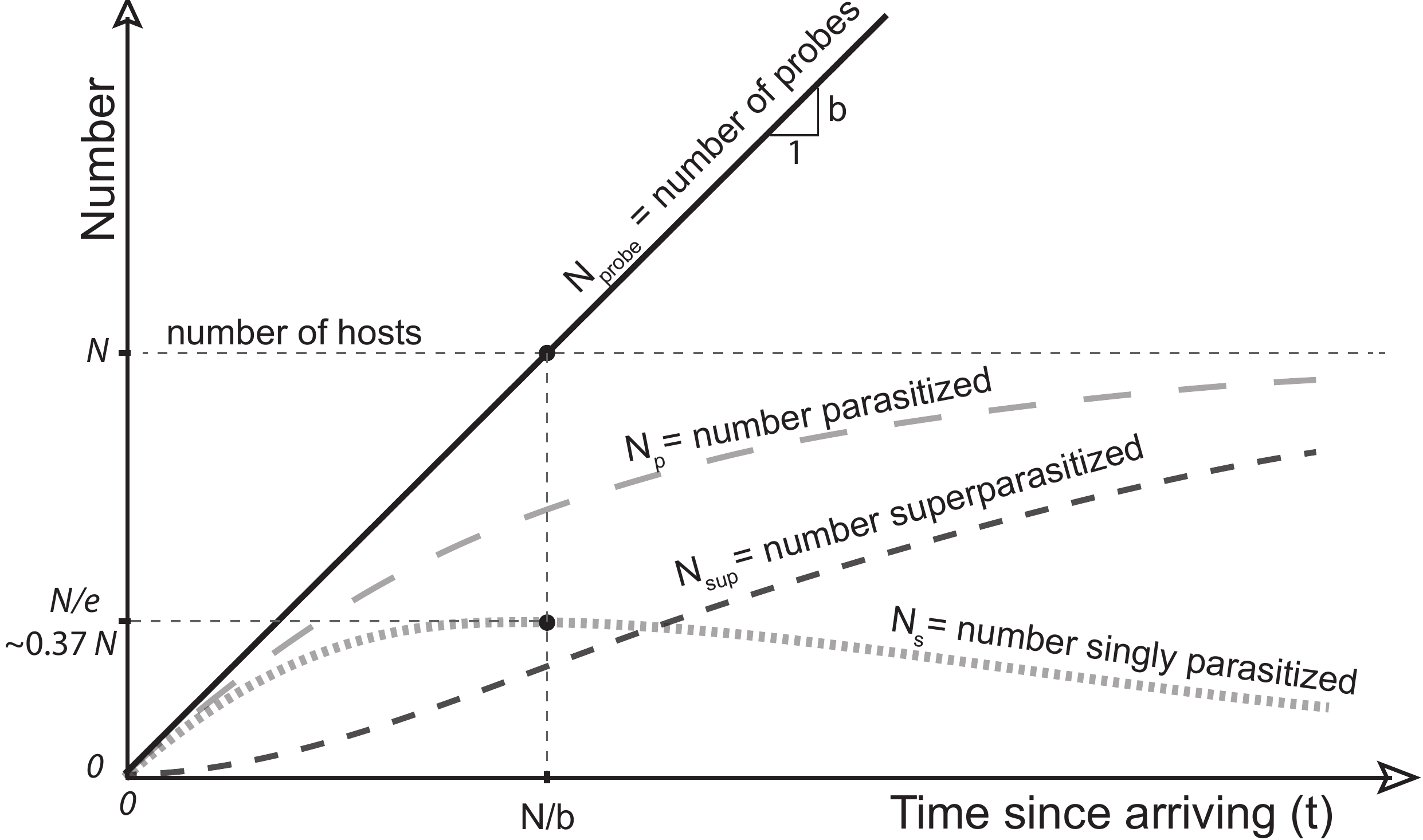} 
\caption{ \baselineskip=2\baselineskip\textbf{Schematic model of parasitism with random probing and no avoidance of superparasitism}. 
The lines represent: the number of times a wasp probes eggs in the cluster (solid line);  the total number of hosts parasitized at least once (long-dashed line);  the number of superparasitized hosts  (dashed line); the number of singly parasitized hosts (short-dashed line). Calculations shown in Appendix \ref{App:Superparasitism}. N  represents the total number of eggs in the cluster. On the singly-parasitized curve the point (N/b, N/e) is at the maximum of the $N_s$ curve; $N/e$ shows the maximum number of singly parasitized hosts the wasp can make. For partial superparasitism avoidance (as done by \textit{H. horticola}), the curve would lie between N, and $N_p$, tending toward zero at very long times ($t$).}
\label{fig:ParasitismEfficiency} 
\end{figure}

\begin{figure}[H]
\centering
\includegraphics[trim=0cm .7cm 1.2cm 2cm, clip, width=.75\textwidth]{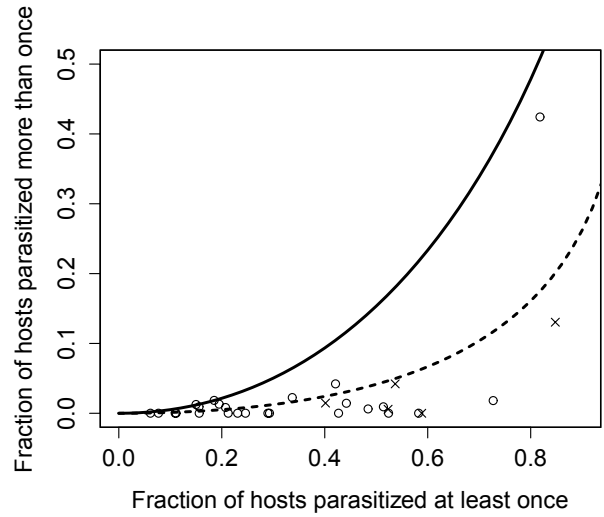} 
\caption{ \baselineskip=2\baselineskip\textbf{Frequency of superparasitism}. 
The fraction of caterpillars parasitized plotted against the fraction containing multiple \textit{H. horticola} eggs, parasitized in the laboratory (circles), and field (crosses). The solid line shows  the expected of fraction of hosts containing multiple parasitoid eggs if wasps choose eggs randomly without avoiding superparasitism. The dotted line shows the best-fit line for the data ($77$\% avoidance of superparasitism). Calculations shown in Appendix \ref{App:Superparasitism},
where the fitted parameter ($z$) is the expected probability of detecting a previous parasitism and not laying an egg (here $z=0.767, SE=0.036, p<0.001$).}
\label{fig:Superparasitism} 
\end{figure}

\begin{figure}[H]
\centering
\includegraphics[trim=.5cm 0.3cm 0.2cm 0.4cm, clip, width=.90\textwidth]{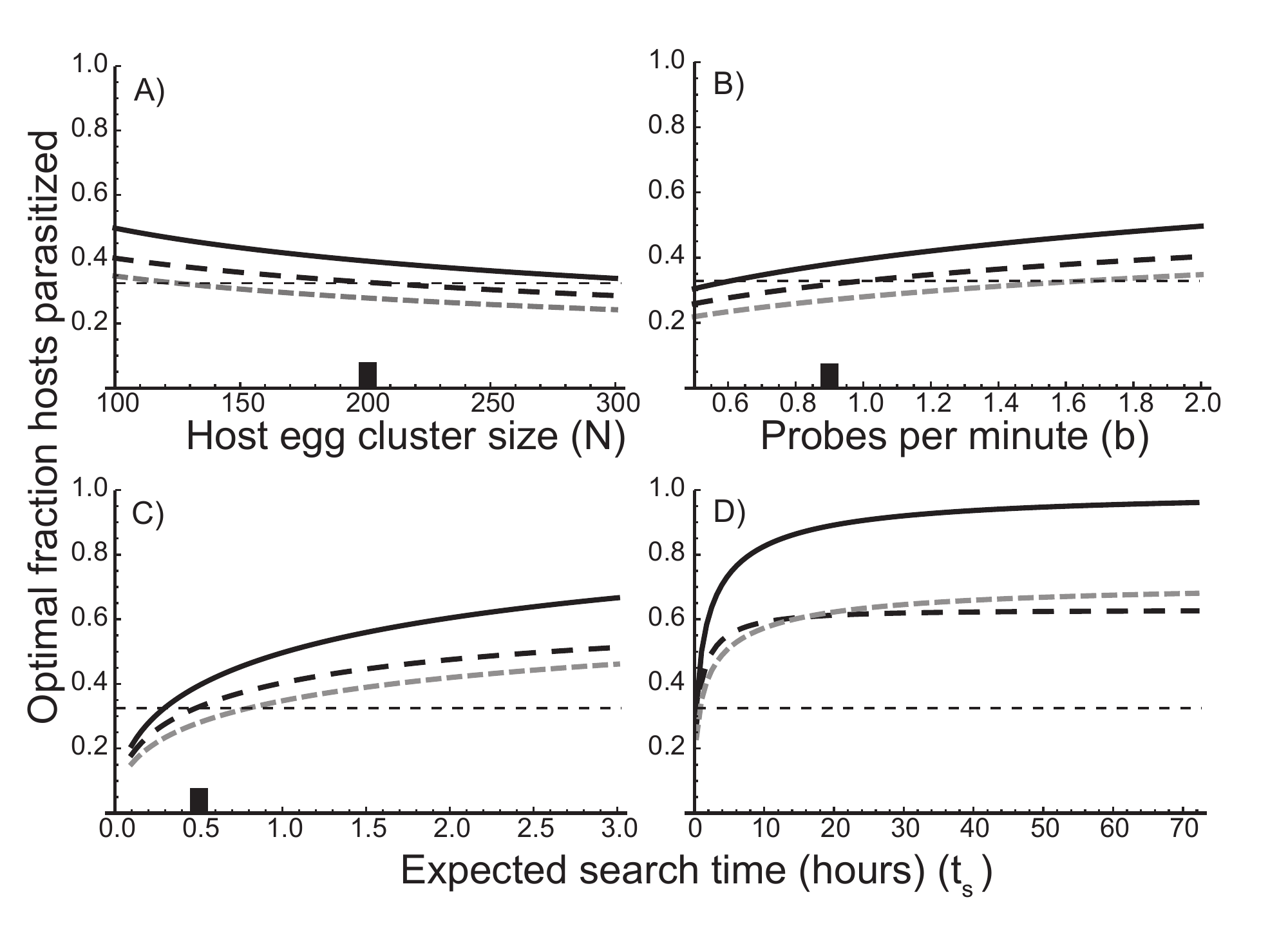} 
 \caption{ \baselineskip=2\baselineskip\textbf{Optimal foraging predictions for the parasitic rate.} Three cases are compared to the observed mean parasitism rate in the field (thin dashed line): parasitized hosts produce one wasp (model $w_1(t)$ (Eqn. \ref{eqn:OptFor})) (bold solid line); assuming parasitoids in multiply parasitized hosts die (model $w_2(t)$ (Eqn. \ref{eqn:OptFor2})) (bold dashed line); and including the cost of hyperparasitism (model $w_3(t)$ (Eqn. \ref{eqn:OptFor3})) (dashed grey line).  For each panel one variable was varied and the rest were held constant (black rectangles) at $N= 200$ eggs \citep{Kuussaari_etal2004}; $b= 0.9$ (Appendix \ref{App:CalcB}), $t_s= 0.5$ hours (best guess for transit time).  Predictions depend on A) N; B) $b$; and  $t_s$ the search time to the next egg cluster at short and long time scales C) and D).}
\label{fig:OFSolns} 
\end{figure}

\begin{figure}[H]
\begin{center}
\includegraphics[trim=0cm .4cm 0cm 0.8cm, clip, width=1\textwidth]{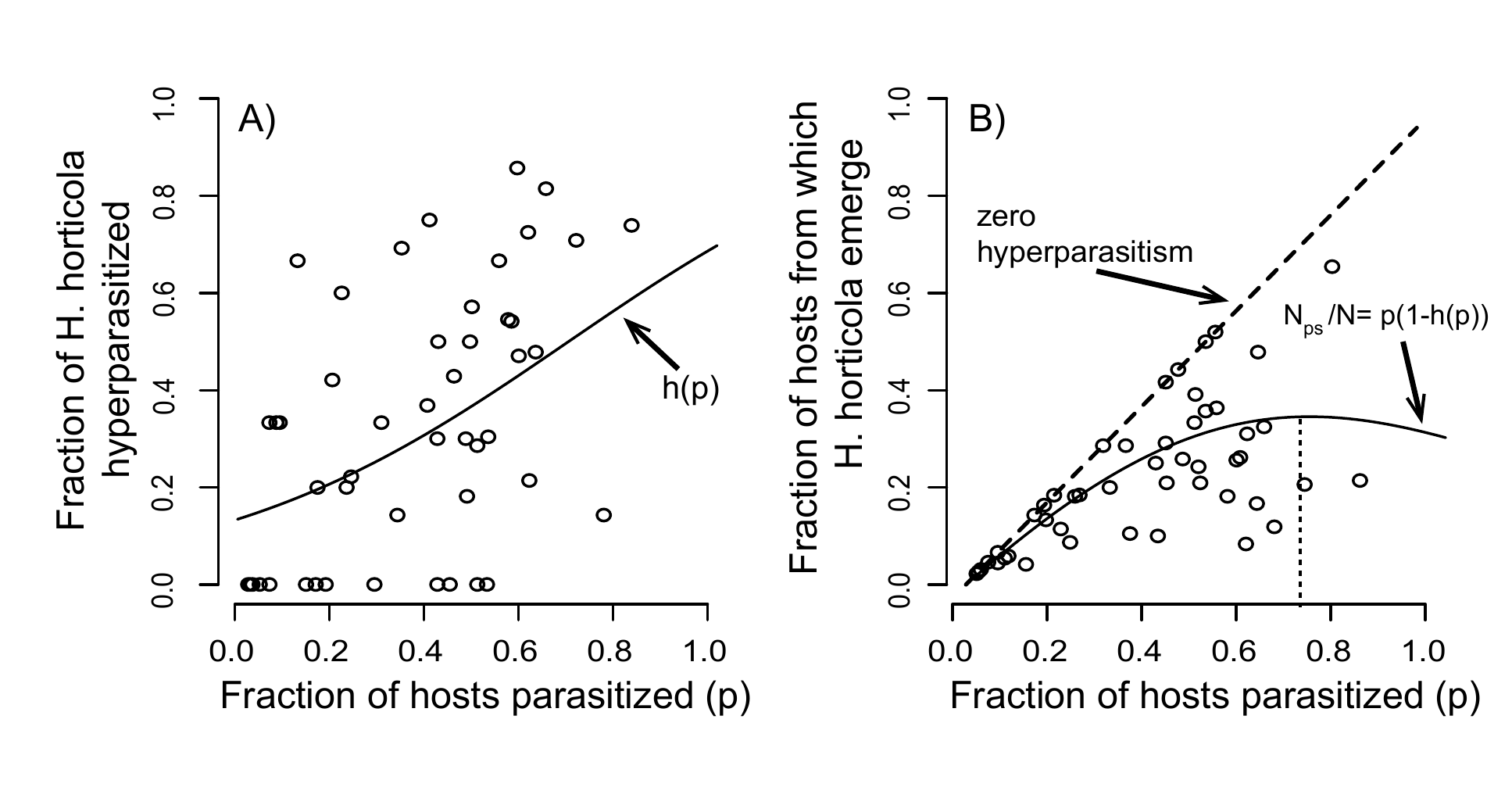} 
\end{center}
\caption{ \baselineskip=2\baselineskip\textbf{Benefits of parasitism under hyperparasitism pressure}. The relationship between the fraction of hosts parasitized by \textit{H. horticola} and A) the fraction of those hyperparasitized; and B) the fraction of hosts from which \textit{H. horticola} eventually emerge.  The dashed line ($y=x$) represents the expected fraction of hosts which produce adult \textit{H. horticola} in absence of hyperparasitism. The solid line shows the best fit curve from logistic regression (h(p)=$1/(1+e^{1.86-2.7p}$) (Eqn. \ref{eqn:h(p)}). Note that taking into account hyperparasitism there is no gain for \textit{H. horticola}  to parasitize at a rate over 0.7 (dotted vertical line in panel B). }
\label{fig:MesData} 
\end{figure}
 
\bibliography{Hyposoter}

\appendix

\section{General Experimental procedures}
\label{App:GeneralExpProc}

Unless noted otherwise, the hosts used in experiments came from a laboratory population of \textit{M. cinxia} maintained in Finland, as described in Couchoux and van Nouhuys (2014). \textit{Hyposoter horticola} were obtained by placing unparasitized hosts in natural populations in  \AA land to be parasitized. After parasitism they were brought back into the laboratory and reared under the same conditions as the unparasitized hosts. 

Upon adults emergence, female \textit{H. horticola} were maintained in the laboratory and fed honey water (3:1). The adult butterflies were also fed honey water (3:1) and placed in cages (3 females + 8 non-sibling males) for one day to mate. After mating, two female butterflies were put in a cage with a host plant (\textit{Veronica spicata}) to lay eggs. When an egg cluster was laid, the plant with the egg cluster was stored until the eggs were close to susceptible to parasitism. Depending on the experiment, they were then exposed to parasitism in the laboratory or placed in a habitat patch in the field to be parasitized by \textit{H. horticola}. 

For the comparison of rate of parasitism in populations from \AA land vs. Morocco (with and without a history of parasitism, respectively) and \AA land vs. Estonia (with and without a history of hyperparasitism, respectively), host caterpillars were collected from each locality. Nine nests were collected from both the Moroccan highlands and \AA land in the autumn 2011 and kept in diapause under laboratory conditions until spring 2012. In the spring of 2012, we collected 11 post-diapause \textit{M. cinxia} nests from Paldiski, Estonia. We then reared all the caterpillars in the laboratory until they pupated. This produced adult \textit{M. cinxia} from \AA land, Morocco and Estonia, and adult \textit{H. horticola} from \AA land and Estonia. 

To obtain host egg clusters from all three origins, \textit{M. cinxia} butterflies from each region were then allowed to mate and oviposit in the laboratory, as explained above. Each female was mated to a non-sibling male from its own origin. For each trial of the experiment, a female wasp was put in a 40 by 40 by 50 cm cage containing a plant (\textit{V. spicata}) with a susceptible egg cluster on it. The host plant species \textit{V. spicata} is common to all three collection sites. In order to reduce variation of behavior, each individual was given parasitism experience before being used in an experiment. A different wasp was used for each observation, and we observed each  parasitism from when the wasp started to probe the egg cluster until it flew off the plant.  This lasted from 10 to 90 minutes.  For details see  Couchoux and van Nouhuys (2014). Afterward we moved the egg cluster to a Petri dish and waited one to three days for the host eggs to hatch. Host caterpillars were then dissected to determine the parasitism rate in each egg cluster. In total 64 egg clusters were parasitized. See table \ref{table:AlandEstoniaMorocco} for a summary of number of replicates for each treatment and results.

\section{Measuring the fitness cost of living in a highly parasitized host nest}
\label{App:Cooperativebenefits}

Parasitized and unparasitized caterpillars from \AA land were obtained as described in Appendix \ref{App:GeneralExpProc}. To assess the effects of rate of parasitism on the performance of \textit{H. horticola} in pre-diapause caterpillars (instars one to five), we put newly hatched caterpillars in 40 composite replicated groups of 40 caterpillars.  We made a well-distributed range of parasitism frequencies by mixing caterpillars from field parasitized clusters with caterpillars from the same laboratory origin that had not been exposed to parasitism. We made aggregate groups of unparasitized caterpillars left undiluted, those mixed 1:1 with caterpillars from nests exposed to parasitism, or entirely from caterpillars from field parasitized nests. We could not create nests parasitized at extremely high rates ($>0.60$).  Young parasitized and unparasitized \textit{M. cinxia} caterpillars are indistinguishable from the outside so we did not know the actual fraction parasitized within each constructed nest until the end of the experiment. Caterpillars developed in these groups under laboratory conditions, and built their silken winter nests.  To assess the quantity of winter silk, we sorted the groups of caterpillars (blind to the level of parasitism) into five groups based on the amount of silk produced. Then we weighed the caterpillars and dissected them to determine which individuals were parasitized.

To assess the effects of parasitism rate on parasitoid performance in post-diapause caterpillars  we used a second set of 37 laboratory-reared and field-parasitized composite groups. We obtained the caterpillars as described above, and reared them until diapause in their original family groups.  After breaking diapause, we mixed families of caterpillars to avoid differences between families, and then put them  composite groups as described above. We then measured days until pupation, and the butterfly and wasp pupal weights. Some caterpillars died in the last instar due to a viral infection. We  dissected these to determine  whether they were parasitized or not.

We analyzed the association of pre-diapause and post-diapause growth rates, weight at diapause and weight at pupation for hosts and parasitoids separately using standard least squares ANOVA with a REML  approach \citep{JMP_2012}. The explanatory variable was rate of parasitism of the group, and group ID was included as a random effect. We analyzed the association of mortality with rate of parasitism using logistic regression with individual survival (0/1) modeled as a function of rate of parasitism of the group and group ID.  Finally,  the association of rate of parasitism with amount of silk produced was analyzed using ANOVA with silk production (level one to five) as an explanatory class variable and group ID as a random effect.

\section{Modeling how reliably wasps avoid superparasitism}
\label{App:Superparasitism}

We assume that at each probe the wasp randomly chooses a host egg from all the eggs in the cluster, and it probes in each cluster many times.  Because about $30\%$ of each roughly 200 egg host cluster is parasitized the wasp must probe, on average, more than $60$ eggs per cluster.  We use the Poisson probability distribution, assuming the number of probes is sufficiently large, to estimate the number of probes a host egg will receive. The probability that the number of times ($n$) a particular host is probed $k$ times is
\begin{align}
P(n=k) &= \frac{\lambda^k}{k!}e^{-\lambda},
 \label{eq:ProbabilityDist}
 \end{align}
where $\lambda$ is the mean number of probes per host, or the total number of times the wasp probes the cluster divided by the number of eggs in the cluster, $\lambda = N_{\rm probe}/N=bt/N$, where $t$ is the time the wasp spends at the cluster and $b$ is the number of eggs probed per minute. 

\paragraph{No avoidance of superparasitism.} 
If every probe by the wasp results in an egg being laid, then the expected numbers of unparasitized hosts $N_{np}$ and singly parasitized hosts $N_s$ can be calculated using equation \ref{eq:ProbabilityDist}: 
\begin{align}
 E[N_{\rm np}] &= N \cdot P(n=0) = \   N e^{-bt/N} \\  
% E[N_{\rm p}]  & =   N- E(N_{\rm np}) \approx N\left(1 - e^{-bt/N}\right)\\
 E[N_s] &= N \cdot P(n=1)  = \ bte^{-bt/N}.
%E[N_{sup}] &= E[N_{\rm p}]-E[N_s]\approx N \left(1 - (1+bt/N)e^{-bt/N}\right).
 \label{eq:noavoidance}
 \end{align}
 These expected values can be used to compute the expected number of parasitized $N_p =   N- E[N_{\rm np}]$, and superparasitized $N_{sup}= E[N_{\rm p}]-E[N_s]$ hosts.

The resulting parasitism as a function of time is shown in Fig. \ref{fig:ParasitismEfficiency}. At given time $t$ the plot shows the numbers of probes $N_{\rm probe}$, parasitized hosts $N_p$, singly parasitized hosts $N_s$, singly probed hosts $N_1$ and superparasitized hosts $N_{sup}$.  As time goes on the expected number $N_{\rm np}$ of unparasitized hosts tends to zero and $N_{\rm p}\rightarrow N$. The number $N_1$ of singly probed hosts increases with time, then decreases towards zero with a maximum of $N_{1}= N/e\approx 0.37(N)$ which occurs when $N_{\rm probe}=N$ at $t = N/b$. 

\paragraph{Avoidance of superparasitism.} 
We continue to assume that the wasp lays eggs in any unparasitized host. But now we assume that superparasitism can be avoided by the detection of a previously deposited parasitoid egg while probing.  The expected fraction of the cluster that remains unparasitized is $p_0=E[N_{\rm np}]/N$. If the wasp  detects prior parasitism with probability $z$ and does not lay an egg when prior parasitism is detected, then the probability that a host egg is parasitized {\em only once}  is 
\begin{align}
p_1&=P(n=1)+zP(n=2)+z^2P(n=3)+\ldots\nonumber\\
&\approx  \lambda e^{-\lambda}+z\frac{\lambda^2}{2!}e^{-\lambda}+z^2 \frac{\lambda^3}{3!}e^{-\lambda}+\ldots\nonumber\\
&= \frac{ e^{-\lambda}}{z}\sum_{j=1}^\infty \frac{(z\lambda)^j}{j!}=\frac{ e^{-\lambda}}{z} (e^{z\lambda}-1).
\end{align}
The probability that an egg is multiply parasitized (superparasitized) is $p_{\rm m}= 1- p_0 - p_1.$

Using data (Fig.~\ref{fig:Superparasitism}) we estimate the probability of avoiding multiparasitism ($z$), by maximizing the multinomial log-likelihood for the model ($p_0(\lambda)$, $p_1(\lambda,z)$, $p_m(\lambda,z)$) given observations on the counts of unparasitized ($p_0$), singly parasitized ($p_1$), and multiply parasitized hosts ($p_m$) in each cluster. The data were fit using  the mle function from the Stats4 package in the statistical package R \citep{R}. We find that $z$ is significantly different from zero ($p<0.0001$). The estimate is $z=0.767$, with $SE=0.036$, that is, we estimate that the wasps detect previous parasitism approximately $77\%$ of the time (dotted line in Fig. \ref{fig:Superparasitism}).

\section{Probing efficiency $b$}
\label{App:CalcB}

We assume that an individual wasp, at an egg cluster of $N$ hosts, probes hosts at a constant rate of $b$ probes per unit time.  After time $t$ the total number of probes is $N_{\rm probe} = bt$.  At each probe we assume that the probability that a given host will be probed is $1/N$, independently of whether it was probed before or not. Once probed, we assume a given host egg is parasitized. If, while probing, the wasp detects previous parasitism,  it may or may not withhold from superparasitizing, as discussed in Appendix \ref{App:Superparasitism}. 

Because we cannot clearly identify a probing event visually (cannot distinguish it from general exploration with the ovipositor), we estimate the probing rate $b$ from the duration of time $t$ spent at a host cluster and the number of hosts parasitized (counted by dissection) $ N_{\rm p}$. With random probing the probability of a host having been probed is  $1-p_0=1-e^{-\lambda}=1-e^{-N_{\rm probe}/N}$ (using Eqn.~\ref{eq:ProbabilityDist} in Appendix \ref{App:Superparasitism}).

\paragraph{Experiment.} We observed  \textit{H. horticola} from \AA land probing and ovipositing into 36 host egg clusters in the laboratory.  These were the same egg clusters used to determine which developmental stages of the eggs are parasitized. The total time spent probing the eggs ($t$) was recorded for each cluster. After all $N$ caterpillars in a cluster hatched we dissected and counted the number parasitized $N_{\rm p}$. We performed logistic regression using a GLM with a binomial error function and logit link function in the statistical package R \citep{R}. We fit the data to the model for $N_p$ (Appendix C) and estimate $b=0.96$ eggs per minute ($p<0.001$),  with  95\% CI of $0.81<b<1.12$.

\end{document}